# In-situ direct visualization of irradiated e-beam patterns on unprocessed resists using atomic force microscopy

H. Koop,[1] D. Schnurbusch,[2] M. Müller,[2] T. Gründl,[2] M. Zech,[1] M.-C. Amann,[2] K. Karrai,[1] and A.W. Holleitner[2,a]

1) attocube systems AG, Koeniginstrasse 11a RGB, 805310 Munich, Germany

2) Walter Schottky Institut und Physik-Department, Technische Universität München, Am Coulombwall 3, D-85748 Garching, Germany

We introduce an in-situ characterization method of resists used for e-beam lithography. The technique is based on the application of an atomic force microscope which is directly mounted below the cathode of an electron-beam lithography system. We demonstrate that patterns irradiated by the e-beam can be efficiently visualized and analyzed in surface topography directly after the e-beam exposure. This in-situ analysis takes place without any development or baking steps, and gives access to the chemical (or latent) image of the irradiated resist.

a) Author to whom correspondence should be addressed. Electronic mail: holleitner@wsi.tum.de

Pushing the size of e-beam lithography (e-BL)-patterns to the few nanometer regime opens up new challenges for both the lithography apparatus and the e-beam resist.[1-5] In order to achieve the best e-BL-resolution, a detailed understanding and control of the e-beam induced processes in the resists is necessary.[6-13] Irradiated regions of e-beam resists exhibit several changes in the physical properties as a result of e-beam irradiation. In poly-methylmethacrylate (PMMA), e-beam exposure gives rise to by-products in form of oxygen and carbon, which can escape the resist during the irradiation.[14,15] In turn, the irradiated film shrinks relative to the unexposed resist, and the physical compression gives access to the chemical (or latent) image of the irradiated PMMA.

At a moderate dose, e-beam irradiation breaks the main chain bonds of PMMA, and the molecular weight is reduced.[16,17] Hereby, irradiated PMMA is more soluble in a developer, enabling a positive tone resist. At a high e-beam dose, a cross-linking process can increase the local density of PMMA, enabling a negative tone resist.[18,19] The cross-linking gives also rise to a physical compression. Therefore, for both a moderate and a high e-beam dose, the latent image can be visualized and characterized as a topographic image by an atomic force microscope (AFM).[18,20-22] Other factors that determine the shrinkage include beam energy, resist thickness, the type of substrate, and the geometry of the irradiated pattern.[17]

So far, latent images have only been characterized by an AFM after bringing the exposed PMMA patterns to ambient conditions.[18,20-22] There, however, adsorption of water and subsequent swelling of the resist can change the chemical properties of the irradiated resists.[15,21,22] Here, we introduce an in-situ AFM characterization method of latent images in e-beam resists. The AFM is based on a tuning fork force sensor[23] and is fully integrated into the vacuum chamber of an e-BL-system. After finishing the exposure process, its effect on the resist is immediately evaluated by imaging the topography of the irradiated resist using the AFM in non-contact mode. Such an in-situ analysis has important implications for the application of chemically amplified resists to high-resolution e-BL. We demonstrate that for PMMA the beam dose can be detected with a sensitivity of $6.5 \pm 0.2$ µC/cm$^2$. Furthermore, the impact of the pattern geometry on the latent image can be directly explored in-situ. In addition, the granularity and homogeneity of the resist are simultaneously characterized. All parameters are essential to optimize the resolution of the e-BL.[6-13]

As depicted in Figure 1(a), the compact and un-obstructive attocube-AFM III is mounted directly below the cathode of an e-BL-system.[24] The relative position of the sample with respect to the e-beam is controlled by piezo positioner[24] blocks A and B (dashed and dotted arrows). The coarse distance of the AFM tip to the sample is further controlled by piezo positioner[24] C [arrow in Figure 1(a)]. In this arrangement the sample can be irradiated by the e-BL-system and independently imaged in-situ with the AFM.

The oscillation amplitude u of the tuning fork at the tip can be described using an effective harmonic oscillator equation[23] $\partial^2 u/\partial t^2 + \gamma \partial u/\partial t + \omega_0^2 u = \omega_0^2 u_0 \sin(\omega t)$, with $\omega_0 = 2\pi f_0$ its natural resonance at frequency $f_0$, $\gamma$ its damping rate and $u_0$ the dither amplitude excitation driven at frequency $\omega = 2\pi f$. When the tip engages into a proximal interaction of force gradient $\nabla F$ with the PMMA surface [Figure 1(b)], the tuning fork resonance shifts to $f_0' \cong f_0(1-\nabla F/K)^{1/2}$ while its phase changes accordingly to $\phi = \arctan\{\gamma f/[f_0^2(1-$ $\nabla F/K) - f^2 ]\}$. We assumed $\gamma \ll \omega_0$. In our measurement $f_0 = 31.9$ kHz, and the resonance full width at half maximum is FWHM = 22 Hz corresponding to a quality factor $Q = f_0/\text{FWHM} = 1450$ and a damping rate[23] $\gamma = 2\pi/\sqrt{3}\text{FWHM} = 80$ sec$^{-1}$. The tuning fork was driven on resonance to have oscillation amplitude (at tip) equivalent to 600 times its natural Brownian fluctuation at 300 K making the tip oscillation amplitude approximately 240 pm. In order to image the sample topography the tip sample distance was regulated at a constant tip amplitude reduction of 10 %. The scan velocity was 1.3 µm/sec corresponding to a typical pixel time acquisition of 10 ms.

Two PMMA resists[25] A2 and A6 are spin-coated onto SiO$_2$/Si wafers. The thickness of the SiO$_2$-layer is 160 nm, and after spin coating the resists have a thickness of 60 nm (A2) and 360 nm (A6).[25-27] Using the e-BL-system Raith e-LiNE,[28] gratings are irradiated into the resists at a constant acceleration voltage of 21 keV and an e-beam dose in the range of 90 to 400 µC/cm$^2$.

Figure 2(a) and (b) show phase and topography images of the resist A2 directly after the e-beam exposure. The (un)exposed areas I (II) of the grating have a width of 1 µm (2 µm). The exposed areas show a shrinkage of about 0.46 nm compared to unexposed areas on the resist [see two top curves in Figure 2(c)]. We would like to note the following points. First, the phase image in Figure 2(a) exhibits the same value for areas I and II. Only at the transitions between the two areas, there is a small reversible phase jump. The constant phase value for areas I and II of the PMMA demonstrate that a topographic change between the two areas dominate the AFM-signal. If an (electrostatic) force in one of the areas was to influence the tip-sample interaction, a constant phase offset between the areas I and II would show up.[23] Second, the noise limit of the AFM apparatus in the vertical direction can be estimated by subtracting two neighboring topographic line cuts as shown in the bottom curve of Figure 2(c). This estimate yields a rms-value of $z_{rms} \approx 40$ pm. This small value shows that the height fluctuations in the two top curves of Figure 2(c) originate from the granularity of the PMMA [triangles]. It is well known that PMMA exhibits grains with a size larger than 50 nm.[29] To first order, the granularity in our PMMA samples stays constant before and after e-beam exposure.

Figures 3(a) to (d) depict in-situ AFM images of the exposed resist A2 recorded after exposure to increasing electron doses. It is apparent that higher e-beam doses give rise to a larger shrinkage of the exposed areas with respect to the non-exposed areas. The probability $\rho(z)$ of having a pixel at a topographical height z is plotted in histograms for each corresponding image in Figures 3(e) to (h). The two peaks appearing in each histogram $\rho(z)$ are perfectly well fitted by two Gaussian $\rho(z) = \rho_b + \rho_e \exp[(z-z_e)^2/2\sigma_e^2] + \rho_0 \exp[(z-z_0)^2/2\sigma_0^2]$, riding on a constant background $\rho_b$ peaking at $\rho_e$ and $\rho_0$ at the location of $z_0$ and $z_e$ the average topography heights on the non-exposed and exposed regions respectively. Here, $\sigma_0$ and $\sigma_e$ are the corresponding standard deviations from $z_0$ and $z_e$.

We extract the shrinkage $\Delta z = z_0 - z_e$ for all exposed gratings with a total standard deviation $\sigma$ in the order of $\sigma = \sigma_0 + \sigma_e \approx 2$ pm. Figure 4(a) shows the shrinkage $\Delta z$ of the resists A6 (black dots) and A2 (white dots) as a function of the e-beam dose D. We can approximate both dependencies with a linear function $\Delta z = \alpha D$, with slopes $\alpha$ of (6.11 ± 0.18) pm/ µCcm$^{-2}$ (A6) and (2.43 ± 0.05) pm/ µCcm$^{-2}$ (A2).

The knowledge of the slope α for a specific resist allows us to judge the quality and the real value of the absorbed e-beam dose in-situ. This can be done by the following estimate: $D \approx z_{rms}/\alpha$ = 6.5 ± 0.2 µCcm$^{-2}$ (for A6) and D ≈ 16.5 ± 0.3 µCcm$^{-2}$ (for A2). The gradient α, however, needs to be calibrated with respect to other parameters, such as the beam energy, the resist thickness, the type of substrate, and the geometry of the irradiated pattern.[17] The latter is demonstrated for resist A6 in Figure 4(b). Generally, Δz depends on the spacing of the gratings. For the data in Figure 4(b), the gratings in the resist A6 have a width ratio of exposed and unexposed areas of 1:1. In Figure 4(b), the Δz data can be fitted with a linear function extrapolating to the origin. In other words, for smaller feature sizes, the unexposed area absorbs almost the same energy per unit area as the exposed area, although this area was not exposed on purpose. This proximity effect has recently been discussed for latent images taken ex-situ at ambient conditions.[20] There, it was estimated that in gratings with a line width of 40 nm, the proximity effect is ~30 %. We have also performed ex-situ AFM measurements on the set of gratings analyzed in Figure 4(b). We detect that Δz increases by a factor of 1.44 ± 0.18 for A6, when the samples are stored at ambient conditions for 3 days [data not shown]. It is noteworthy, that this increase of Δz can be caused by both an increased shrinkage of the exposed areas or an increased height of the non-exposed areas. Both effects are irrelevant for the vacuum in-situ analysis.

To conclude, the presented in-situ visualization and characterization technique has crucial prediction capabilities thanks to the linear dependence of resist shrinkage on exposure dose which we revealed in this work. It allows investigating the quality of the exposure without any resist processing, allowing a step-wise proximity correction and re-exposing the desired spots with the e-beam. Compared to the traditional approach of imaging the post-processed resist for instance in a scanning electron microscope, the in-situ AFM delivers more physically valuable information, such as polymer grain size and distribution. Furthermore, the resulting line edge roughness of irradiated pattern can be investigated directly after exposure, excluding the impact of chemical processes. Therefore, the in-situ AFM has great potential to be used as a classification and verification tool for the development of new high resolution e-beam resists and for the exposure of really small structures.

We thank P. Weiser for great technical help. Furthermore, we gratefully acknowledge financial support by the DFG (Ho 3324/4), the German excellence initiative via the "Nanosystems Initiative Munich (NIM)" as well as the "Center for NanoScience" (CeNS) in Munich.

**FIG. 1.** (Color online) (a) Drawing of the atomic force microscope (AFM) incorporated into the vacuum chamber of an e-beam lithography (E-BL) system. The footprint of the microscope is about 15 x 33 mm. (b) The AFM-tip exhibits a tuning fork which is scanned across the surface of a sample by the help of piezo positioners A, B, and C. The topography of the PMMA is read out by the shift of the frequency of the tuning fork.

**FIG. 2.** In-situ AFM characterization of a grating in A2-PMMA directly after e-beam exposure. The (un)exposed areas I (II) of the grating have a width of 1 µm (2 µm) (dose: 190 µC/cm$^2$). (a) Phase and (b) topographic map of the irradiated area. (c) Two adjacent line sweeps of (b) are shown (top curves). The lines are presented with an off-set of z = 0.6 nm for better visualization. Bottom curve: experimentally determined root-mean square (rms) value $z_{rms}$ of about 40 pm.

**FIG. 3.** (a) to (d) show AFM images of 1µm line / 2µm space pattern irradiated in resist A2 at a dose of 90, 140, 190 and 290 µC/cm$^2$. Plots (e) to (h) show the respective height histograms (data points) of the AFM images of (a) to (d). The lines are fitting curves involving two Gaussian peaks as discussed in the text.

**FIG. 4.** (a) Resist thickness shrinkage after exposure plotted as function of exposure dose D. Resist A2 (open circle) and A6 (closed circle). (b) Thickness shrinkage for grating with a line:space ratio of 1:1 as a function of the line width for resist A6 at constant dose of 190 µC/cm$^2$. The linear fits are empirical.

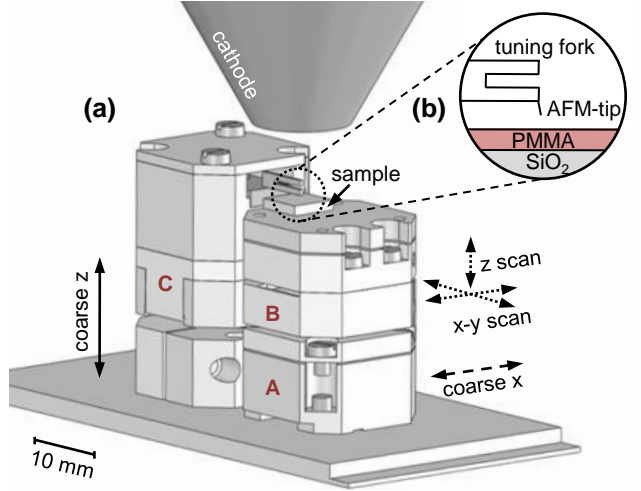

FIG. 1.

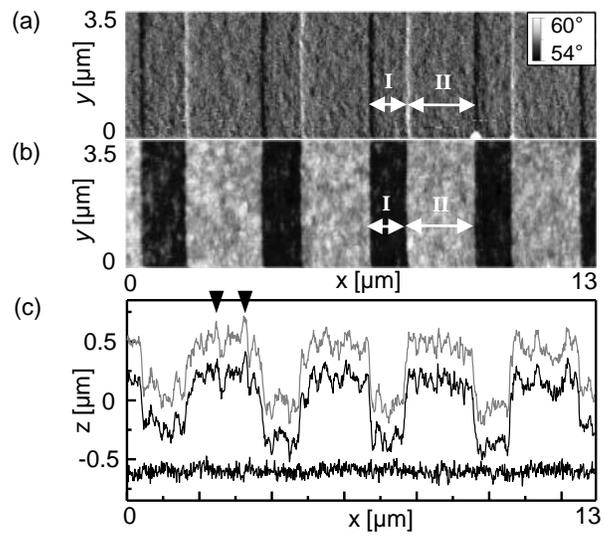

FIG. 2.

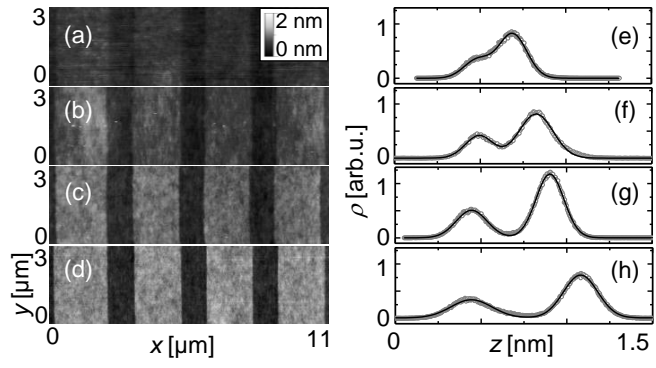

FIG. 3.

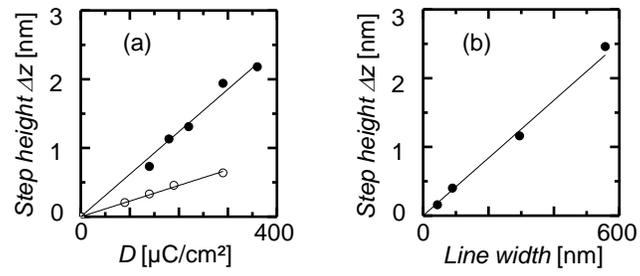

FIG. 4.